# A new 2-D Model to analyze uncertainty sources of sparse sea surface $CO_2$ partial pressure

**Xiang Li[1,3], Minhan Dai[1,2] and Guizhi Wang[1,2]**

Correspondence to: M. Dai, mdai@xmu.edu.cn

1.State Key Laboratory of Marine Environmental Science, Xiamen University, Xiamen, China, 2. College of Ocean and Earth Science, Xiamen University, Xiamen, China, 3. School of Physics, Huazhong University of Science & Technology.

## Abstract

In order to better comprehend the global carbon cycle and predict the prognosis for the response to climate change, accurate assessment of sea-air $CO_2$ flux is necessary. Comparing to the relative homogeneously distribution of atmospheric $CO_2$, the $pCO_2$ in the sea surface water is exposed to huge spatio-temporal variability, which leaves a prominent uncertainty resource. Many regional studies typically divided the observational $pCO_2$ data into grid boxes so as to obtain enough data points statistically for their calculatio. However, using the data inside the grid box areas to represent its holistic property (such as standard deviation to represent spatial variance) will mix up three different uncertainty sources. First, the analytical error in the $pCO_2$ determination and the associated environmental parameters used in deriving $pCO_2$. Second, the spatial variance because of inhomogenous spatial pattern of sea surface $pCO_2$, especially the region with a dramatically dynamic circumstance like: coastal areas, boundaries or fronts and etc. Third, the estimation process in undersampling condition, specifically, this kind of uncertainty origins from the process that using a sparse data to represent its holistic property of the box area. Common uncertainty quantification by Standard Deviation will mix up the different sources of uncertainty. In this paper, it introduces an optimized procedure to determine three sources of uncertainty ($1^{st}$ analytical error, $2^{nd}$ spatial variance, $3^{rd}$ bias from undersampling.) using the combined remote sensing-derived and field-measured pCO2 data. In order to provide a comprehensive error assessment report.

## Special Section
Pacific-Asian Marginal Seas

## Key Points

- Three sources of uncertainty of sparse $pCO_2$ data.
- A new 2-D model introduced in Kriging Estimation.
- Coupled with remote sensing-derived data to fit the 2-D model.
- The uncertainties of sparse $pCO_2$ data are estimated.

# 1. Introduction:

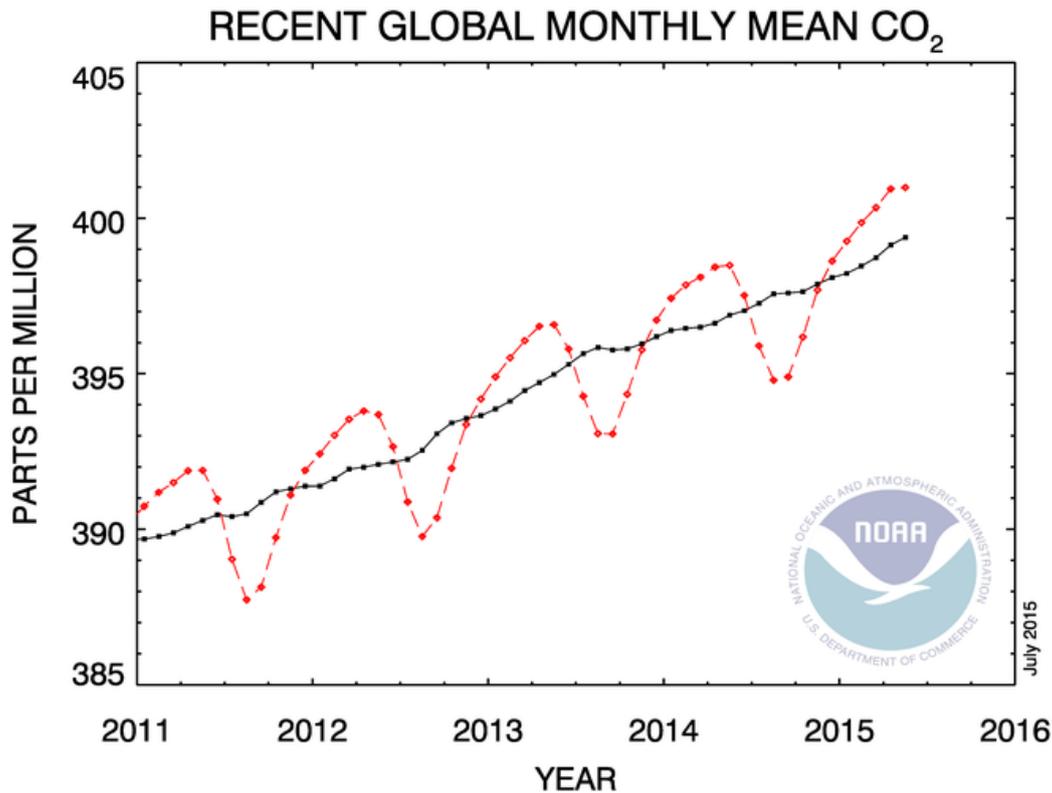

Figure.1 From Global Carbon Budget 2014

The continuing rapid accumulation of $CO_2$ in atmosphere inevitably reach to an unprecedented level in human history, the 400 ppm mark [Global Carbon Budget 2014]. In order to better comprehend the global carbon cycle and predict the prognosis for the response to climate change, accurate assessment of sea-air $CO_2$ flux is necessary [Takahashi et al., 2009, 2002, 1997]. Since 1993, a global cooperated program, the CDIAC Ocean Carbon Data Management Project, has been established for collecting discrete and underway measurement $pCO_2$ data. Thanks to tremendous effort contributed by individual investigators and groups, up till 2014, more than 9.0 million measurements of surface water $pCO_2$ data points made over the global oceans have been included in the LDEO database [Version 2013]. However, comparing to the relative homogeneously distribution of atmospheric $CO_2$, the $pCO_2$ in the sea surface water is exposed to huge spatio-temporal variability, which leaves a prominent uncertainty resource. It directly casts a shadow on the reliability of estimating the sea-air flux.

Many regional studies typically divided the observational $pCO_2$ data into grid boxes so as to obtain enough data points statistically for their calculation[e.g., Takahashi et al., 2009; Zhai et al., 2013]. However, using the data inside the grid box areas to represent its holistic property (such as standard deviation to represent spatial variance) will mix up three different uncertainty sources [Wang et al., 2014]. First, the analytical error in the $pCO_2$ determination and the associated environmental parameters used in deriving $pCO_2$. Second, the spatial variance because of inhomogenous spatial pattern of sea surface $pCO_2$, especially the region with a dramatically dynamic circumstance like: coastal areas, boundaries or fronts and etc [Sweeney et al., 2013]. Third, the estimation process in undersampling condition, specifically, this kind of uncertainty origins from the process that using a sparse data to represent its holistic property of the box area.

Because of inhomogeneously and sparsely distribution of sampling station points is common, correctly and efficiently quantifying those three kinds of uncertainties is strongly required. This study bases on the spatial analysis and Kriging estimation, using a new two-dimensional spatial correlation model to quantify the contribution of the spatial variance and undersampling uncertainty in sea surface $pCO_2$ data and further to optimize the assessment of sea-air $CO_2$ flux.
1. Analytical Error (Em)
2. Spatial Variance
3. The risk from estimation in undersampling condition.

Given that its necessity when evaluating different studies, our approach shall have widely applications as an uncertainty analysis tools.

## 2. $pCO_2$ database and methodology to quantify the uncertainty of sparse data.

Monitoring the variability of sea surface $pCO_2$, people use ships and other platforms generates large amounts of data from heterogenous sources. Since 1993, a global cooperated program via a number of U.S. and international ocean-observing programs, the CDIAC Ocean Carbon Data Management Project has been established for collection of discrete and underway measurements $pCO_2$ data. The database used in this study are from an underway measurements in ECS August, 2009 and Global Ocean Surface Water Partial Pressure of $CO_2$ Database (Version 2013).

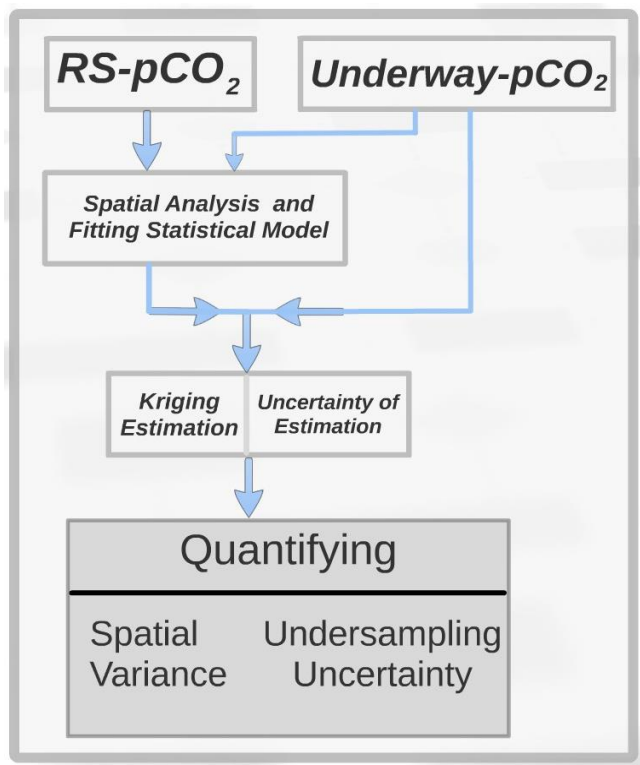

Figure.2 Flowchart of the quantifying uncertainty for $pCO_2$ data. RS-$pCO_2$ represents remote sensing-derived $pCO_2$ data, and underway represents field-measured $pCO_2$ data.

For a given sea surface $pCO_2$ field data set, the analytical error, Em, is normally calculated by aggregating all the errors introduced in the measurement and data reduction of $pCO_2$ in many studies[]. In this study, it will focus on a standard procedure for estimating the rest two sources of uncertainty, the spatial variance and undersamping. And this methodology will be more efficient with aid of a concurrent high-spatial-resolution satellite-derived $pCO_2$ data set from [Bai et al., 2015] in the same studying region. But, it is not required so. The protocol of this method will be summarized in a simplified way in Figure.2. First, using spatial data (Underway or Remote-Sensing measurements) to do the spatial analysis for estimating a new two-dimensional spatial correlation model. Second, combining the obtained statistical model and underway measurements to quantify the correlation of each data pair. Then, basing on Ordinary Kriging (OK) Estimation to provide a full coverage data set of the study region. Finally, estimate the spatial variance via calculating the standard deviation of the full coverage data set and the undersampling uncertainty from the estimation variance of OK.

The details will be as the following steps.

## 2.1 Spatial analysis and circular segment model.

For a given spatial data set, such as ESC August 2009, normally they have an inherent property, the spatial dependence in attribute values, which means the values for the same attribute measured at locations that are near to one another tend to be similar, and tend to be more similar than values separated by larger distances. In this study, in order to provide an efficient and simple algorithm, we assume the dependency structure is also isotropic, the same on both axes, which is an ideal condition. Although this assumption does not sufficiently consider the ocean system, which is persistently dynamic in physical, chemical and biological process, it reflects some basic information or property from data itself. The particular steps to quantifying the structure of spatial dependency in a data set will be fully discussed in [Robert Haining, 1993].

$$\hat{C}(h) = [(1/N(h))\underbrace{\sum_i \sum_j (z(i)z(j))}_{[(i,j)|d(i,j)=h\pm\Delta)]}] - \bar{z}(i)\bar{z}(j) \quad (1)$$

Where, C(h) is the estimate of spatial covariance at distance h. Figure.2 is a conceptual example of a typical semi-variogram for the case where spatial dependence in attribute values. This spatial autocorrelation analysis is the basis for the next estimation.

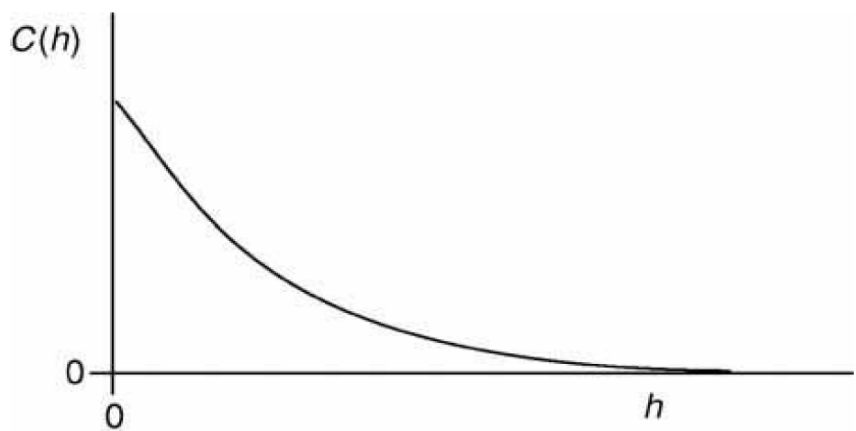

Figure.3 (Conceptual model for an autocovariance function C(h)). [i]

From the above spatial correlation analysis, the next step is fitting statistical model. Considering dealing with a 2-dimensional problem, we proposed a new 2-D statistical model as shown in Figure.3. This model represents the correlation of two data points by calculating the segment area of them, also in the case of an isotropic weakly stationary processes. This means that any permissable covariance function also can be used as a model for the semi-variogram of a weakly stationary process, the detailed explanation are given by [Cressie, 1991, p87].

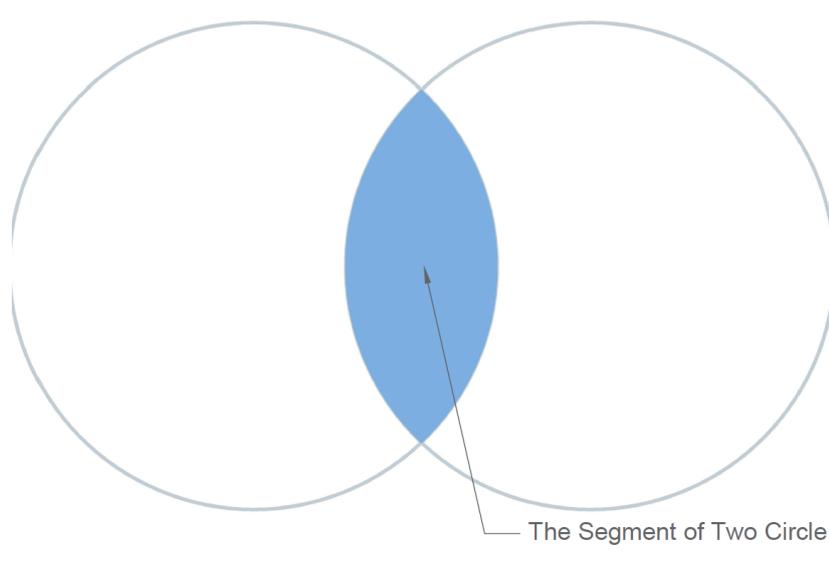

Figure.4 The 2-dimensional covariance statistical model.

Where the overlap of two circulars at a distance d will be calculated by the flowing formula:

$$A_{overlaping} = \pi r^2 - d\sqrt{r^2 - d^2/4} - 2r^2 \arcsin \frac{d}{2r} \quad (d <= 2r) \quad (2)$$

The covariance function of this 2D model is:

Circular Model:
$$C(h) = \begin{cases} C_0 + C & h = 0 \\ \frac{C}{\pi r^2}[\pi r^2 - d\sqrt{r^2 - d^2/4} - 2r^2 \arcsin \frac{d}{2r}] & 0 < h \leq a \\ 0 & h > a \end{cases} \quad (3)$$

The sill is the plateau a function reaches and which corresponds to C(0). The range is the distance at which the plateau is reached. Detail see in [Robert Haining, 1993, p295].

## 2.2 Deriving from Kriging Estimation.

After finishing the spatial analysis and model fitting, the three constants of statistical model, Circular Model, will be known. And using this obtained model, we could statistically quantifying each data pair's correlation on average.

Then, deriving the equations from ordinary kriging with two conditions in terms of 2D covariance function. This prediction is the homogeneously linear combination of the data set that minizizes the mean squared prediction error and will be fully discussed in [Cressie, 1991].

$$\begin{vmatrix} C_{11} & \cdots & C_{1n} & 1 \\ \vdots & & \vdots & \vdots \\ C_{n1} & \cdots & C_{nn} & 1 \\ 1 & \cdots & 1 & 0 \end{vmatrix} \begin{vmatrix} \lambda_1 \\ \vdots \\ \lambda_n \\ \mu \end{vmatrix} = \begin{vmatrix} C_{01} \\ \vdots \\ C_{0n} \\ 1 \end{vmatrix} \qquad (4)$$

The above equation set is to calculate a single estimation point, derived by us following Kriging ordinary method. And the results will be the following:

$$Z_0^* = \sum_{i=1}^{n} \lambda_i Z(x_i) \qquad (5)$$

Estimation of a single point Z0.

$$\sigma_{OK}^2 = C(0) - \sum_{i=1}^{n} \lambda_i C(x_0, x_i) + \mu \qquad (6)$$

Uncertainty of OK estimation that brings in.

Using this method, provide a full coverage estimation of sparse data set. Then, using

$$\sigma_s^2 = \frac{1}{N} \sum_{i=1}^{N} (Z_i - \bar{Z}), \qquad (7)$$

to estimate the spatial variance via calculating the standard deviation of the full coverage data set and the undersampling uncertainty from the estimation variance of OK.

$$\sigma_u^2 = C(0) - \sum_{i=1}^{n} \lambda_i C(x_0, x_i) + \mu \qquad (8)$$

Finally, integrating the three uncertainty, the analytical error, the spatial variance and the undersampling variance, and in this case, three of them all are totally independent. The total uncertainty () is:

$$\sigma_T = \sqrt{E_m^2 + \sigma_s^2 + \sigma_u^2} \qquad (9)$$

## 3. Applications of the Uncertainty Quantification to different cases.

From above, we introduced a modified Kriging estimation method. This method of quantifying uncertainty was applied to different case studies, such as, the underway surface $p$CO$_2$ data from the East China Sea collected in August 2009, shown in Figure . The standard procedure will be like, spatial analysis (Figure.4), estimation of 2-D statistical model and estimation of full coverage data (Fig.5.a) and its uncertainty (Fig.5.b).

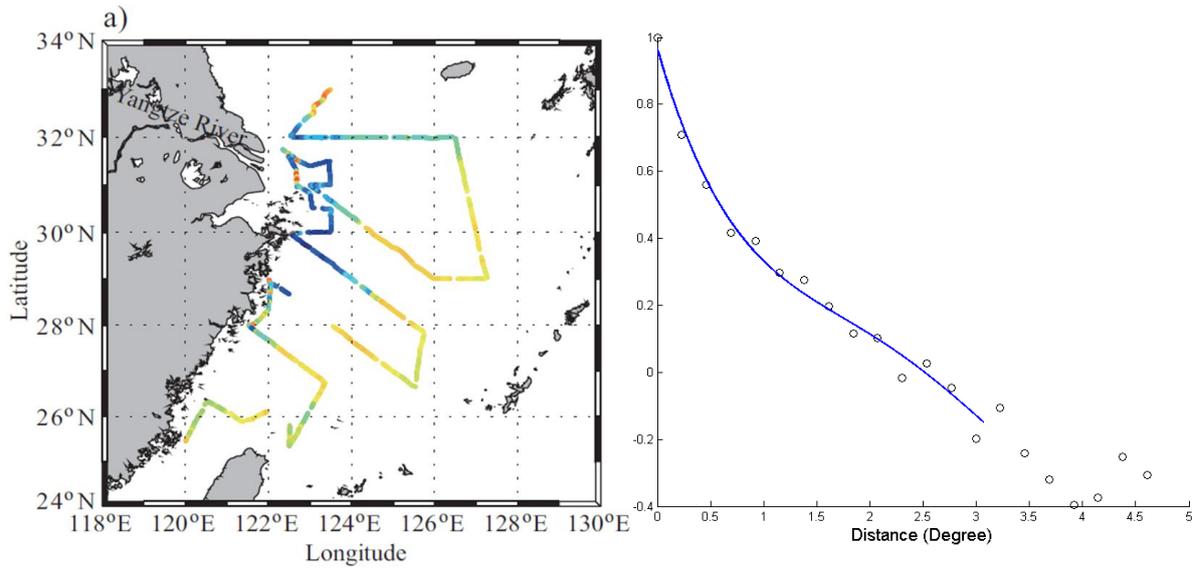

Figure.4 (a) Underway surface $pCO_2$ data from the East China Sea collected in August 2009 from [Wang et al., 2014 ], (b)Spatial covariance function to this data set.

## 3.2 Results Validation (ESC August, 2009)

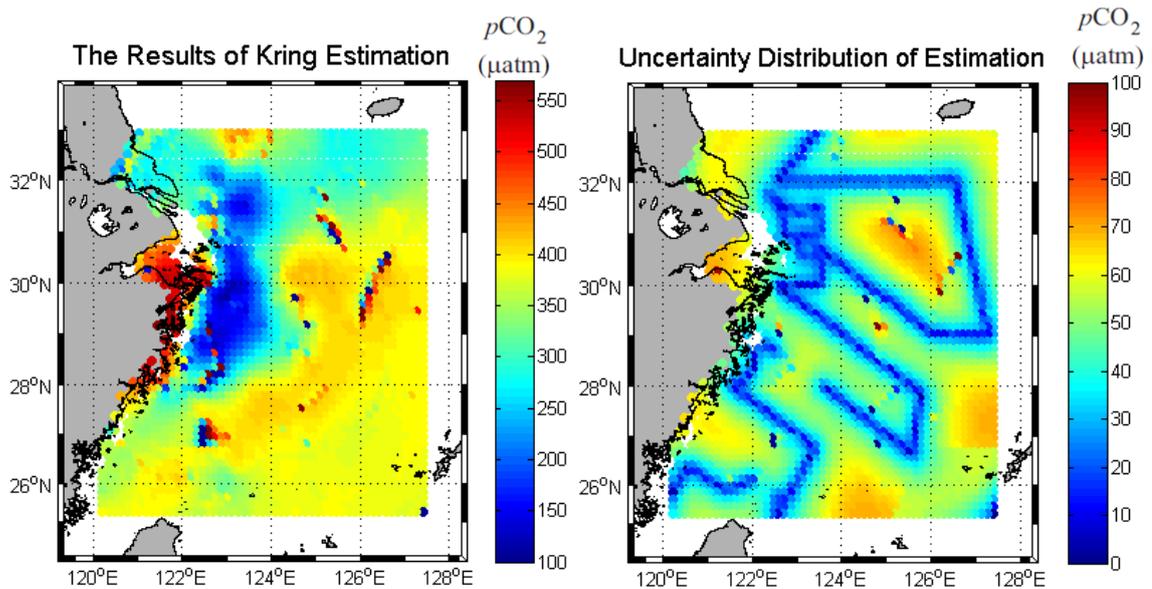

Figure.5 (a) A full coverage estimation of ESC August, 2009 data set. (b) The undersampling uncertainty distribution of each estimation points.

While calculating this data set or other case studies of a field trip. There are always some locations of data stations having the same coordinates, which means more than one measurement at a single location. And it will cast a critical problem for this OK method, because the equation set () is singular, not full rank, so that they can not be solved. In this study, we have used a trick to deal with this problem, randomly displacing each data station a very little bit in space, that the displacement is tenth lower the minimum resolution of underway measurement. However, it is still need more studies to evaluate how much uncertainty will be taken in due to this rough tactic.

Meanwhile, we have a concurrent high-spatial-resolution satellite-derived $pCO_2$ data set from [Bai et al., 2015] in the same studying region shown in Figure.6. We can see a prominent coherence between estimations results (Figure.5.a) and Remote-Sensing data (Figure.6). And most importantly, the distribution of undersampling uncertainty was strongly proportional to the density of underway data, which is quite decent and logically. Here, in this method, it uses the sum of the squared deviation, N times variance, as the maxim of the estimation uncertainty. And for each estimation points, the uncertainty will decrease when having some adjacent data points. Basing on correlated data and information will reduce the risks of estimation, or specifically, uncertainty.

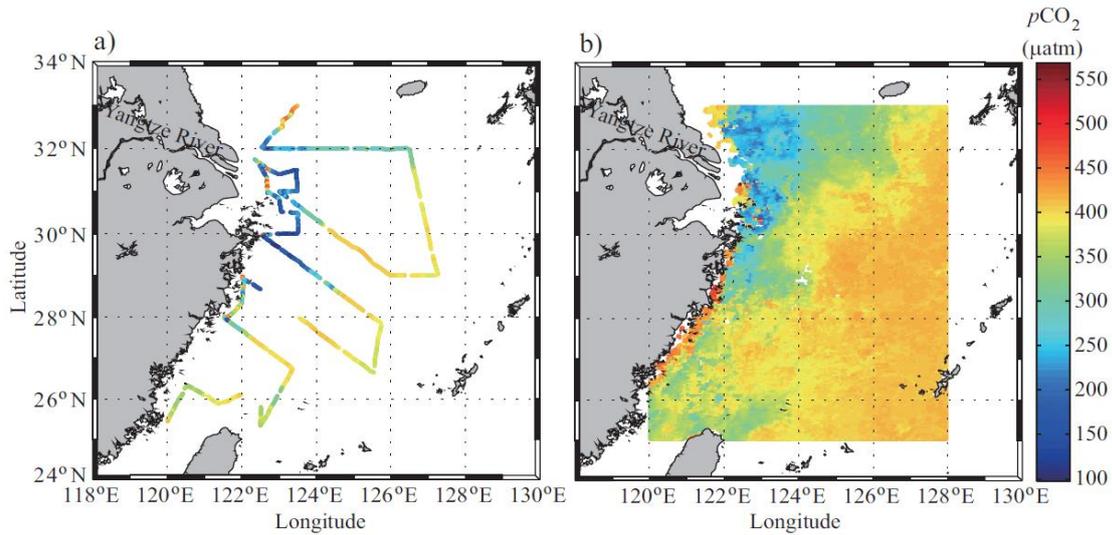

Figure.6

### 3.3.1 Other Case Studies- If Heterogeneity?

From the current huge amount of $pCO_2$ database, Aiming to avoid the largely temporal variability, we only picked up some field trips within a short time period, normally less than half a month. Two typical cases will be discussed in the following.

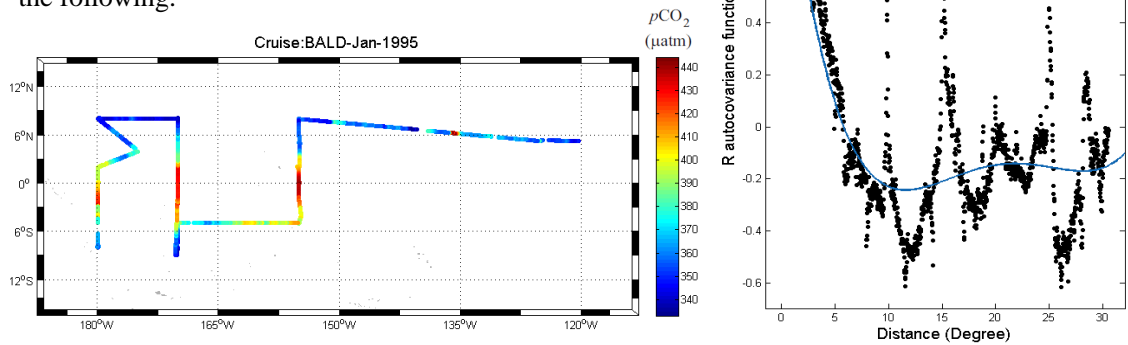

Figure.7 (a).Underway measurement of Cruise: Bald Jan 1995, data from CDIAC. (b) The autocovariance function of this data set.

From above, it showed an inconformity between the auto-correlation function and ideal model. So as the result, it showed a poor quality of not only the full coverage of estimation but also the uncertainty of estimation points. So, to promote the quality of results, we recommend to geographically define subsets of data set if heterogeneity is suspected as the underway measurement showed, the observation coverage of this field trip can not fully detect its spatial structure.

### 3.3.2 If Overlapping?

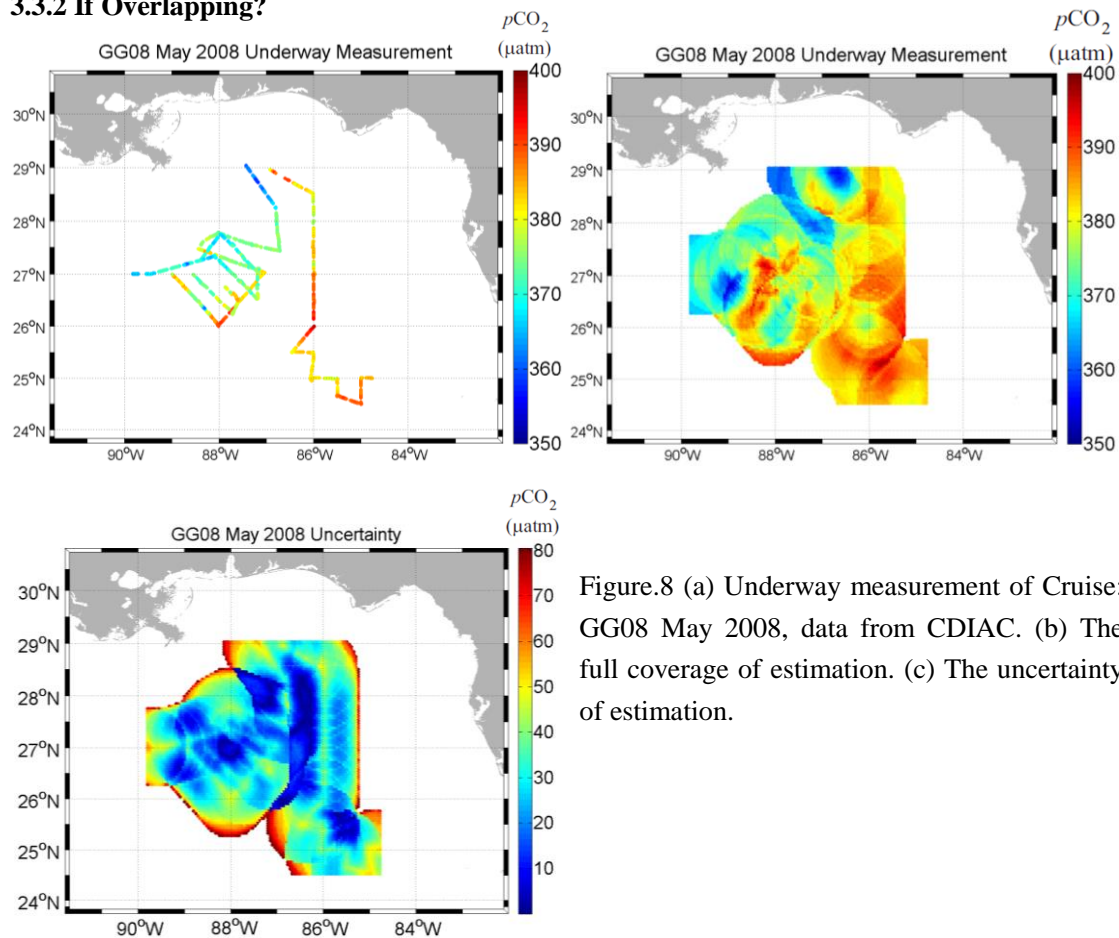

Figure.8 (a) Underway measurement of Cruise: GG08 May 2008, data from CDIAC. (b) The full coverage of estimation. (c) The uncertainty of estimation.

As shown in Figure.8, here is an another typical status with conformity between the auto-correlation function and ideal model. But, some data stations have overlapped. This will result in a phenomenon called nugget effect and also weaken the efficiency of this method.

### 4. Summary

From all above, we introduced a general method to calculate the three uncertainties of spatial sparse data, such as our focus, the sea surface $pCO_2$ data. It successfully provides a full coverage of estimation and its uncertainty for calculating the spatial variance and undersampling uncertainty of studying region. And if with aid of a concurrent high-spatial-resolution satellite-derived $pCO_2$ data set, this method will provide a more robust result. However, the application of this method to some field trip measurement from CDIAC database still reveal some critical weakness in several typical conditions, for instance, heterogeneity and overlapping. Considering this study will have a widespread application in the future, there are still some optimizing work need to be done.